# How Interaction Designers Use Tools to Manage Ideas

NANNA INIE, Aarhus University
PETER DALSGAARD, Aarhus University

This paper presents a grounded theory-analysis based on a qualitative study of professional interaction designers (n=20) with a focus on how they use tools to manage design ideas. Idea management can be understood as a subcategory of the field Personal Information Management, which includes the activities around the capture, organization, retrieval, and use of information. Idea management pertains then to the management and use of *ideas* as part of creative activities. The paper identifies tool-supported idea management strategies and needs of professional interaction designers, and discusses the context and consequences of these strategies. Based on our analysis, we identify a conceptual framework of ten strategies which are supported by tools: *saving*, *externalizing*, *advancing*, *exploring*, *archiving*, *clustering*, *extracting*, *browsing*, *verifying*, and *collaborating*. Finally, we discuss how this framework can be used to characterize and analyze existing and novel idea management tools.

---

## 1 INTRODUCTION

The fields of HCI and interaction design are well-known for studying impacts of novel interfaces and tools, and perhaps less known for studying the impact of tools that are already used in professional practice (Smith et al. 2009; Pedersen at al. 2018; Dalsgaard 2017). In the words of Stolterman, this often leads to research outcomes that are difficult to apply in practice, because the research is based on an inadequate understanding of how design happens in professional settings (Stolterman 2008). Therefore, discovering, analyzing, and discussing how professionals already use tools is critical for the disciplines of interaction design and HCI research. In this paper, we investigate the question: how do professional interaction designers use tools to manage ideas, and how might tools be characterized in terms of the way they are used in practice, rather than by their functional properties?

Interaction designers face various challenges when working with tools, among these are organizational barriers, the ever-changing state of technology, as well as technical difficulties (Dow et al. 2006). They work with digital materials on a daily basis, and they employ an extremely versatile assemblage of digital and analog tools throughout their career. Interaction designers have a broad array of analog and digital tools available to aid their idea management, from sticky notes and notebooks to software applications such as Evernote and Trello. Tools are distinguished by their ability to support and extend the limited cognitive systems and abilities of humans, and previous research has shown promising results in their support of creative cognition (Smith et al. 2009).

## 1.1 Motivation and positioning

The framework presented in this paper has been developed to assist researchers and tool developers in understanding the nature and influence of idea management tools in professional design work. The framework is intended to support and inspire researchers within interaction design and HCI in *analyzing*, *characterizing*, and *evaluating* any given digital or analog tool in terms of the role the tool plays in professional idea management. Idea management is just one of many aspects of interaction design practice, and this paper contributes to the discussion of how to support creative practices in professional settings. Creativity support tools can be generally defined as tools which support and extend their users' ability to make 'creative discoveries', whether they support the stages of gathering information, generating hypotheses, imagining and refining solutions, or disseminating and validating concepts (Shneiderman 2009). Current research in tool development does not provide many guidelines for how to design creativity support tools, according to Shneiderman; however, the potential is immense: *"Creativity support tools enable discovery and innovation on a broader scale than ever before; eager novices are performing like seasoned masters and the grandmasters are producing startling results. [...] The risks are high and the scientific methods novel, but the payoffs are substantial in bringing about thrilling moments of scientific discovery and engineering innovation"* (Ibid.).

We use the following descriptions of the general term 'tool':

*Platform*: Platforms can 'host' or encompass several other tools. Examples would be 'computer', 'smartphone', or 'pen and paper' (the latter of which can in practice mean everything from sticky notes to sketchbooks). We use the term to indicate on which platform a tool is used when the tool is cross-platform, for instance Evernote or Notes, which can be used on both phone, laptop, and tablet platforms.

*Tool:* When we speak of 'tools', we mean a tool which serves a well-defined purpose for the designer, and may run on several different platforms. Examples would be different software applications (Evernote, OneNote etc.) or variations of analog tools (notebook, sticky notes, whiteboard etc.).

*Auxiliary tool*: A fine-grained description of tools, more commonly used by researchers than practitioners (those in our studies, at least). This definition is sometimes used by researchers to describe: *"individual tools within design applications such as Adobe Illustrator and InDesign. [...] individual panels and commands such as color pickers, alignment commands, levels panel, Adobe Photoshop filters, etc."* (Maudet 2017).

Our research is primarily focused on *tools*, because that description corresponds to our study participants' understanding of the word. A body of work has previously explored how we might develop auxiliary digital tools to function across different software applications, which we find an interesting line of research (e.g. (Ciolfi Felice et al. 2016; Jalal et al. 2015; Maudet 2017; Maudet et al. 2017)), but which we do not explore in this paper.

In this paper, we systematically analyze how and why professional interaction designers use tools to manage design ideas. The paper offers three contributions: 1. It provides descriptive insights into how professional designers use existing tools to manage ideas. Idea management is a fundamental practice for all creative practitioners, but one that is rarely given deliberate research attention (Inie et al. 2018a). These insights contribute to a stronger, fundamental understanding of how interaction design happens in professional settings. 2. The paper analyzes and synthesizes these insights into a grounded theory of professional idea management. 3. The paper offers a conceptual framework of strategies for tool-support of idea management. The framework is intended to help researchers and developers of creativity support tools characterize and analyze existing and novel idea management tools.

The paper is based on an in-depth empirical study: a series of interviews with 20 professional interaction designers. During the interviews we had the opportunity to ask the designers to demonstrate how their idea archives were structured, what was in them, and how they used these archives on a day-to-day basis. We use the term 'interaction designer' to describe someone who ideates for and/or gives form to interactive products, environments, systems, and services (Cooper et al. 2007) with careful attention to forming or transforming the user experience (Forlizzi and Ford 2000). The distinctions between industrial job descriptions such as 'interaction designer', 'UX designer' and 'experience designer' are often unspecified, but from a research perspective, designers in these professions share numerous skills and characteristics that are significant from the perspective of design idea management. First, interaction designers practice design constantly and over an extended period of time. As we will detail in the background section, capturing, managing, and using design ideas are pivotal skills during a professional design career. Second, their design 'material' (user experience and user interaction) is often difficult to represent and manipulate using existing tools (Löwgren and Stolterman 1998; Dow et al. 2006). How does an interaction designer represent experiential concepts like '*feeling overwhelmed*' or '*sleek performance*'? And third, interaction designers are technically literate. They have the language and, often, reflection ability to communicate their relationship to and choice of tools, because they are trained in describing these things. The work of interaction designers revolves around the use of digital tools and systems, for which reason we can expect them to have devoted thought and effort into creating appropriate and sufficient workflows using both digital and analog tools (an expectation which was confirmed by the study we present in the paper).

Our inquiry and analysis build on the findings of previous studies of idea management for creative professionals (Coughlan and Johnson 2008; Sharmin et al. 2009; Inie and Dalsgaard 2017) and of Personal Information Management for knowledge workers in a broader sense (Boardman and Sasse 2004; Efimova 2009; Kaye et al. 2006; Odom et al. 2012). Our findings differ in the sense that, beyond describing the insights from our empirical studies, we seek to create a conceptual framework which can be used in the analysis and evaluation of other tools. We do not intend to map all available tools for idea management, nor to quantify the most used tools on a global scale (which has been done much more thoroughly in other reports, i.e. (Loop 2018) and (Vinh 2015)), rather, we investigate unique user stories and systems in depth.

We are also very specific about analyzing *use* of tools, rather than functional properties embedded in the tools. This is based on the premise that technologies can be assessed only in their relation to the environments of their production and use (Suchman et al. 1999). If we base a framework on tool functionality and not tool use, we would likely end up with a framework that would be less usable for understanding real-life use practices. Bernal et al. (2015) addressed the need for a change in focus on the needs of *designers,* rather than design *products*: "*Current computational tools are design-centric, with interfaces from the perspective of the physical components, rather than designer-centric, with a focus on supporting the actions that designers execute while they manipulate the patterns that drive the arrangement of the parts*". A decade before that paper, Kidd (1994) expressed a similar pledge: for computers to *inform*, rather than passively storing *information*: "*Computer support for knowledge work might be better targeted on the act of informing rather than on passively filing large quantities of information in a "disembodied" form*". By positioning our research within a discussion of creativity support tools, we hope to assist and inspire other researchers in HCI to analyze and evaluate tools in terms of their active support of the social, creative, and reflective human designer.

## 2 RELATED WORK

### 2.1 What is idea management?

Idea management is a complicated matter to study, especially in a longitudinal perspective. Most knowledge we have about long-term use of idea management tools is based on personal accounts and small-sample studies (Amabile and Mueller 2008; Efimova 2009; Erickson 1996). In their study from 2008, Coughlan and Johnson (2008) categorize the process of idea management into three essential purposes:

- Retention and organization of ideas
- Feedback, evaluation and development of ideas
- Communication and collaboration of and on ideas

Other empirical studies have found similar categories of activities (Sharmin et al. 2009; Inie and Dalsgaard 2017; Inie et al. 2018a). One study highlights an additional main category of idea management: retrieving and reusing ideas (Efimova 2009). This paper describes the author's activities of idea management (using her personal blogging processes as an example) as the following:

- Low-threshold creation of blog entries
- Organizing and maintaining content in a flexible and personally meaningful way
- Retrieving, reusing and analyzing blog entries
- Engaging with others around blog content.

Since there is currently no standard definition of which activities exactly constitute idea management, we propose an overarching theory and set of descriptions to frame the activities involved. This model is derived based on both previous studies as well as the empirical studies of this paper. The model is presented in the findings section.

*2.1.1 Personal Information Management.* Idea management can be said to be a subcategory of Personal Information Management (or PIM), in that 'ideas' is one of many categories or types of information that need managing throughout a professional creative career. We will describe the particularities about 'ideas' as opposed to general information in the next section. Personal Information Management is a term used in HCI to describe the collection, storage, and retrieval of digital and analog information (Jones 2010), such as emails, reference files, copies of finished projects, et cetera. Several in-depth studies have been conducted on how office workers manage information, often involving extensive ethnographic field work, e.g. (Whittaker and Hirschberg 2001; Boardman and Sasse 2004; Barreau and Nardi 1995). The amount of field work carried out in this discipline accounts for most of our existing knowledge about strategies of information management of professionals. A central objective of PIM is to ensure access to the right information, in the right format, and of sufficient completeness and quality to meet the professional's current need (Jones 2004). One of the goals of research in PIM is therefore to define those needs and to develop systems to help users achieve them.

Although technology's advance means we have more tools and systems than ever available to manage our personal information, it does not necessarily lead to increased satisfaction. On the contrary, empirical studies have showed that many professionals experience frustration with their information becoming fragmented (Teevan et al. 2006; Inie et al. 2018a). In 1994 Kidd declared that developers of PIM tools *"have lost sight of humans as highly-tuned learners and actors whose internal form is constantly changing in order to refine their ability to act in the world"* (Kidd 1994). Therefore, Kidd suggests that (as described in the previous section) support tools should devote more attention to actively *informing* the person, rather than serving as a passive storage unit. Thus, to actively inform the interaction designer, we need to discover which informational purposes ideas serve to designers, in order to be able to enhance the computational support. It is the authors' argument that design ideas bears very unique potential, and that this potential should not be ignored in the development of personal information management systems and tools.

*2.1.2    What are ideas?* The concept of ideas is theoretically and historically closely tied to the practice of creativity. For instance, professional creativity is broadly said to refer to the deliberate and continuous generation and development of novel and useful ideas (Kaufmann and Baghetto 2009; Biskjaer et al. 2010). Although creativity is a *desirable* characteristic of design, and exceptional designers are creative thinkers, creativity is not a *necessary* condition for design (Alexiou et al. 2009). Interestingly, design is commonly described as a creative activity: *"... there can be no guarantee that a creative 'event' will occur during a design process (...) However, in every design project creativity can be found"* (Dorst and Cross 2001).

Cognitive science has offered more elaborate analyses of ideas than interaction design research. The Geneplore Model (Finke et al. 1992) describes ideas as *discoveries* formed in the mind on the basis of mental preinventive structures - precursors for the final externalized creative products or ideas. Preinventive structures usually refer to visual patterns, object forms, or mental models (Ibid.), and may, in a design context, also refer to cognitive structures that rely on external support such as sketches or prototypes (Christensen and Schunn 2007). These discoveries are then explored and evaluated against external and internal constraints (such as knowledge about external requirements or internal expectations or personal taste) before they come to a form of expression.

What is characteristic about a *design* idea (in relation to the creativity research-understanding of an idea) is that the design idea is oriented towards moving a design process forward. A design idea can be directed towards framing or reframing the problem statement, discovering an opportunity to work with, suggesting a full solution for the design problem, or part of a solution for the design problem (Inie and Dalsgaard 2017a). Design ideas are essential in practicing creative design. In the next section we will present some of the most prominent research on the importance of idea management for the interaction designer.

*2.1.3    Idea management and the interaction designer.* The goal of interaction design work is to arrive at a concrete or abstract product as a result of conscious actions and decisions by the designer (Biskjaer et al. 2010). Ideas are essential in moving from a problem state to a desired outcome, in that ideas express a design vision, aimed at solving a design problem (Löwgren and Stolterman 2004). In models of the design process, activities of design roughly consist of iterative phases of defining a problem, collecting data, generating ideas, and selecting the most promising idea to move forward with (Howard et al. 2008) (see Fig. 1).

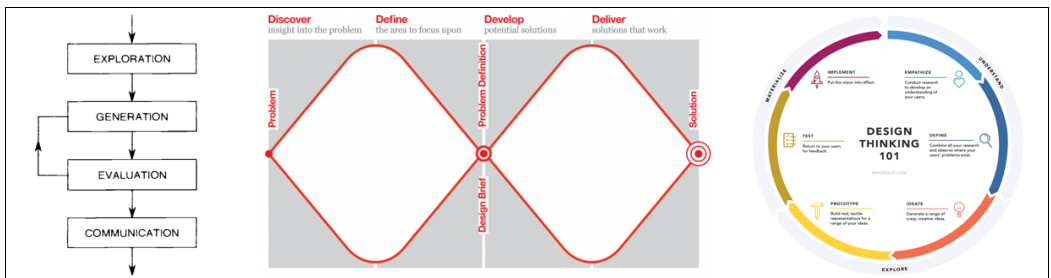

Fig. 1. Some models of the design process. From left to right: A simple four stage model of the design process (Cross and Roy 1989), The Double Diamond Model (Design Council 2006), and The Design Thinking Process (Gibbons 2016).

The ability to generate many or very good ideas and the ability to select the best ideas have been areas of extensive research in the fields of interaction- and engineering design. Many of these studies are aimed towards developing methods for generating more or better ideas (usually termed *idea fluency* and *idea quality*, respectively), for example (Dahl and Moreau 2002; Dix et al. 2006; Girotra et al. 2010; Goldschmidt and Sever 2011; Halskov and Dalsgaard 2006; Howard et al. 2011; Perttula et al. 2006; Siangliulue et al. 2015; Sosa and Dong 2013). These methods are tested in more or less controlled settings, and the methods are seldom used in professional design practice, but instead by students or experiment participants chosen by the researchers. In similar vein, research contributions that investigate the ability to select promising ideas (for instance (Toh and Miller 2015; Nelson and Stolterman 2003; Badke- Schaub and Gehrlicher 2003; Cardoso et al. 2016; Starkey et al. 2016; Shah et al. 2000; Goldschmidt and Tatsa 2005)) often look at discourse during a design process where it is possible to evaluate the outcomes, that is, often a student design course. Such studies indicate that the ability to generate and manage design ideas can be said to be a fundamental competency for the interaction designer.

In professional practice, the value of a design idea is highly dependent on the context it is to be employed in, giving designers a reason to store ideas until they are in a position to use them (Coughlan and Johnson 2008). According to Gaver and Bowers (2012) and Löwgren and Stolterman (2004). The work of a great designer is extensively based on experience from similar design cases - often more than it is based on theoretical knowledge. Buxton (2010) argues that it often takes a decade for a good idea to have practical value in the world, which makes deliberate idea management critical to the interaction designer.

Most professional interaction designers manage externalizations of their ideas in various formats and for several reasons (Kaye et al. 2006). Most often, idea management is guided by improvised ad hoc assemblies of archives and tools, as idea management is not a skill which is specifically taught. Previous studies have found designers to purposefully avoid formalized explication of processes and knowledge, posing very high demands to systems that require them to do so (Shipman and Marshall 1999).

Externalization can be described as the active shaping of the world as an intellectual resource, and it is a core activity for most professional designers (Dix and Gongora 2011). In a more direct sense, externalization can be anything from verbalizing an idea to actively shaping it through interaction with the environment, for instance in the form of modelling or prototyping. In the tradition of cognitive science, externalization is considered any expression of *computational offloading* (Scaife and Rogers 2005) or discoverable manifestation: *"a way of taking information or mental structure generated by an agent and transforming it into epistemically useful structure in the environment. It is a way of materializing structure that first was mental"* (Kirsh 2009). Externalizations of interaction designers are particularly interesting in the scope of idea management, because interaction designers work with "materials without qualities" (Löwgren and Stolterman 2004) - or at least with qualities that can be challenging to create accurate representations of (Dow et al. 2006). As a consequence, numerous tools are in play as 'idea management tools'. In our study we found, perhaps surprisingly, that interaction designers have very few follow-up questions when we ask 'which tools they use to manage ideas'. This indicates that even if the processes are not standardized, idea management is a familiar activity for professional interaction designers.

## 2.2    Tools and the design process

In design research, *tools* often encompass both analog and digital tools. They can be distinguished by their role in aiding the designer's creative cognition (Hollan et al. 2000; Hutchins 1995), rather than by their technical properties: *"[tool] can denote a range of artefacts; and relative, in that it can be any artefact that is employed as a means to transform the situation. That is, an*

*artefact becomes an instrument of inquiry when we use it as such"* (Dalsgaard 2017). It is recognized that tools play an important role in helping designers do much more than just give form to design artifacts; tools support design creativity and exploration by guiding perception and understanding of design problems and solutions, and by helping designers see, understand, explore, and experiment (Ibid.). Tools are parts of co-adaptive phenomena in which they help the designer shape their environment; meanwhile, tools affect human behavior itself (Mackay 1990). In this section we will give an overview of related work of the influence of tools on interaction design practice.

While there is a general understanding and concurrence about the activities involved in creative design work, we have less of a vocabulary to describe how specific tools influence these activities in a frame of creative design. Howard et al. (Howard et al. 2011) describe the following roles of tools in creativity support: a) As a task framing tool during the analysis phase, b) As an idea generation tool during the generation phase, and c) As a selection or evaluation tool in the evaluation phase. However, it is widely recognized that creativity also happens outside idea generation meetings, when the designer is alone or engaging in spare time activities (Amabile and Mueller 2008; Coughlan and Johnson 2008; Finke et al. 1992). Tools that are "phase-specific" therefore do not support longitudinal idea management.

Dalsgaard (Dalsgaard 2017) identified five qualities of tool support which contribute to our understanding of how tools affect professional design practice:

1. **Perception** (tools help designers perceive and understand the design situation and formulate the design problem),
2. **Conception** (tools help designers understand and articulate the problems they face as well as develop hypotheses about how to address these problems),
3. **Externalization** (tools allow designers to make imagined design solutions part of the world),
4. **Knowing-through-action** (tools allow new knowledge to be generated through acting with an instrument), and
5. **Mediation** (tools support mediation between actors and artefacts in a design situation and establish stable shared points of reference).

These five qualities express cognitive attributes that tools support through the design process. As such, they explain properties of existing tools which can be embedded simultaneously in the same tool, rather than isolated in different tools. A 'task framing tool' (Howard et al. 2011), for instance a mood board, can simultaneously support both perception, conception, externalization, and mediation.

While several studies have shown very promising results in developing curation and ideation tools, often integrating analog tools with digital affordances, (i.e. Dorta et al. 2008, Mendels et al. 2011; Lindley et al. 2013, Lupfer et al. 2016), none of these tools have yet been adopted into large-scale use. Some commercial idea curation tools with pervasive success in different markets exist - platforms such Pinterest, Dribbble, and Behance. There has been some research in how Pinterest is used as a curation tool (i.e. Linder et al. 2014; Scolere and Humphreys 2016), although very little on whether such tools are used as part of the creative process, especially in professional contexts.

Previous research in the role of tools has informed this paper in multiple regards. The work presented here is distinguished in two particular aspects: first, it offers a general theory of idea management processes in professional settings. Second, the analysis highlights and details the strategies taken by professional interaction designers to achieve creative goals.

## 3 METHODOLOGY

This paper presents a rich qualitative study based on a method of qualitative, semi-structured interviewing. Qualitative research can be described as enacting a local, action-oriented approach of investigation and applying small-scale theorizing to specific problems in specific situations (Berg et al. 2004). Some of the advantages of this approach are that the method is suitable for *discovering* influences on creativity and for capturing the complexity of organizational creativity. The method is less suitable for determining causal relationships of creativity and other factors, or for generalizing to other individuals (Amabile and Mueller 2008). Qualitative interviewing is an appropriate method for providing insights into *objectives* of interaction designers, rather than only behavior. A relatively small sample allows for thorough engagement with the data in a way where individual differences and distinctive approaches do not get lost in the general (Berg et al. 2004). That is, if a large number of designers use one tool to perform a certain activity, and a single designer uses a different one, it does not necessarily mean the one designer should be ignored. Rather, we might look to the one designer to be inspired to develop new and better systems because they may face the needs of the many ahead of time (von Hippel 1986).

### 3.1 Participants and interview form

We conducted semi-structured qualitative interviews with 20 professional designers, each interview lasting approximately one hour. The interviews were structured in topics corresponding to the activities central to idea management (as described in the background section): capturing ideas, managing ideas, retrieving ideas, and collaborating on ideas. We interweaved factual questions (e.g. "Which tools do you use to...?", "How do you...?") with generative questions (e.g. "Take me back to a time when you...", "Imagine the idea tool for..."). When relevant, we asked to see examples of the designer's ideas. An overview of our interview guide is shown in Table 2.

We interviewed 14 male, and 6 female designers (more information about the participants is shown in Table 3). Participants were recruited via the authors' personal networks, mailing lists, and Facebook groups for UX designers. The age span was between early 20s and late 40s, with experience in design ranging between 1 and 15+ years. We did not choose the designers based on their experience or demographics, but rather based on getting a varied sample of different types of designers. 20 interviews were enough to develop 'saturated' categories of information (Creswell 2013), meaning we could identify patterns of behavior which overlapped in all the categories of our framework.

Table 2. Qualitative interview guide.

| | |
|---|---|
| Introduction | Intro 1: Can I ask you to state your name and affiliation, and that you are OK with this interview being audio recorded and used for research purposes?<br>Intro 2: What is your background in interaction design?<br>Intro 3: What kinds of interaction design do you work with?<br>Intro 4: How much of your time do you work alone vs. together with others when generating and developing ideas? |
| Capturing ideas | 1.1 Which tools do you use to do capture your ideas?<br>• When you're at work?<br>• When you're at home?<br>• When you're at "inconvenient places" (i.e. on a walk, in the shower, at yoga class etc.)?<br>1.2 Can you remember the last time you captured an idea? Describe what happened.<br>1.3 Imagine the ideal tool, in your mind, for continuously capturing ideas.<br>• What would the interface of this tool be like?<br>• What key features would it have?<br>1.4 Why do you capture ideas? What's the end goal-product? And how does archiving contribute to that? |
| **Organizing ideas** | 2.1 Where do you keep your ideas?<br>2.2 How do your ideas look? E.g. sketches, audio files, texts, image collections etc.<br>2.3 Which tools do you use to make them look this way?<br>2.4 Imagine the ideal tool, in your mind, for storing ideas so they are easy to find and use when you need them.<br>• What would the interface of this tool be like?<br>• What key features would it have? |
| **Retrieving ideas** | 4.1 Do you ever look at your old ideas?<br>• Why/why not?<br>4.1.a If yes: How do you use your old ideas for later projects?<br>4.1.b Take me back to the last time you went through an idea archive of yours. What did you learn from it? |
| **Collaborating on ideas** | 3.1 Which tools do you use when you collaborate with others in generating/developing ideas?<br>3.1.a Why these tools?<br>3.2 Do you ever experience difficulty in representing your ideas so you can communicate them to others?<br>• If you think back to the last time you faced this situation: describe what happened?<br>3.3 Imagine the ideal tool, in your mind, for collaborating on ideas with your colleagues or team.<br>• What would the interface of this tool be like?<br>• What key features would it have? |

Because interaction designers have such varied job descriptions and backgrounds (from sociology to computer science), we have chosen not to base our analysis on differences in the designers' current job description. A designer with a background in graphic design may be more inclined to use, for instance, Adobe Illustrator, to develop sketches, but we have no way of distinguishing between whether this would be a difference in training or personal preference. We observed some patterns in preferences for tools, which seemed to correlate more with the designer's training or education than their current job description. The preferences were related to preferred tools for *development* of ideas, however, more than to capture or organization of ideas. For instance, the application Sketch was a recurrently utilized tool for UX designers who designed interactive interfaces, but was not used for e.g. capture of ideas.

Interviews are inherently retrospective, and we are mindful of the fact that we can only report on what participants tell us and how they think or remember to behave. Therefore, we also asked each designer for a "tour" of their idea archives and examples of how their ideas looked. This method worked well to uncover tacit or forgotten knowledge, as it often brought up stories that the designer had not previously shared.

Table 3. Details about interview participants.

|    | Gender, age, job description | Ys of design experience |
|----|------------------------------|-------------------------|
| P1 | Female, late 20s. Works in a large IT-providing company in San Diego, USA. Focus on UX-design on one project. | 9 |
| P2 | Male, mid 30s. Background in Computer Science. Has worked with game design but currently works in academia. Based in San Francisco, USA. | Not specified |
| P3 | Male, early 30s. Game designer at a medium-sized game development company based in Aarhus, Denmark. SCRUM-responsible for his team. | 7 |
| P4 | Male, 40s. CEO of large, world-wide design company. Currently based in New York. Works with design strategy. | 15+ |
| P5 | Male, 40s. Freelance graphic and UX designer. Based in southern Germany. | 15+ |
| P6 | Male, mid 20s. Works as an interaction-/product-/UX designer at a design agency in Mountain View, California, USA. | 2 |
| P7 | Male, early 40s. UX designer at a medium-sized design agency in Aarhus, Denmark. | 7+ |
| P8 | Male, early 20s. Product designer/interaction designer with a focus on software design and experience design. Based in bay area, California, USA. | 1 |
| P9 | Male, 40s. Founder and CEO of medium-sized design company based in San Diego, USA. Background in graphic and web design. | 10+ |
| P10 | Male, 40s. Leading design strategy at a medium-sized design company based in San Diego, USA. Works closely with clients. | 16 |
| P11 | Female, mid 20s. Experience Designer at a software company based in California, USA. Focus on visual design. | 3 |
| P12 | Female, late 20s. UI and strategy designer for an app. Located in Copenhagen, Denmark, but collaborates with USA-branch. Background in visual design. | 7 |
| P13 | Male, 30s. Freelance brand designer and artist. Based in San Diego, USA. Works a lot with space and wayfinding design. | 11 |
| P14 | Female, late 20s. Interaction designer at an IT-provider based in San Diego, USA. Focus on interface design and service design. | 1 |
| P15 | Female, late 20s. UX and UI designer at a large industrial company in Bjerringbro, Denmark. Background in sociology. | 4 |
| P16 | Male, early 30s. Works as a UX designer in a Nordic, digital agency, Aarhus (Denmark) branch. Background in multimedia design. | 6 |
| P17 | Male, mid 20s. UX and product designer (primarily mobile app) for a sharing economy-based startup company located in Copenhagen, Denmark. | 5 |
| P18 | Female, early 40s. UX researcher and designer at Danish, Aarhus-based branch of a large, international e-commerce company. | 5 |
| P19 | Male, early 20s. Intern UX designer at large international design company based in Aarhus Denmark. Studies digital design at university level. | 1 |
| P20 | Male, early 30s. Industrial designer at small electronic product design company based in Aarhus, Denmark. | Not specified |

**Grounded theory-analysis**

The interviews were transcribed and analyzed using a grounded theory approach (Strauss and Corbin 1990; Creswell 2013; Gibbs 2018). We approached the analysis by first dividing the transcriptions into discernable chunks of meaning, and identifying open, descriptive codes (Table 4).

Table 4. Categories identified during open coding.

| Head category | Sub-categories |
|---------------|----------------|
| Idea forms and representations | *To do-lists, visual vs. text, screen dumps, bookmarks, notes, sketches on paper, information, prototypes, talking as prototyping, analog>digital, digital>analog* |
| Software | *Evernote, reminders, Slack, PowerPoint/Keynote, Adobe Photoshop/Illustrator, Asana, Google Keep, Pinterest, tool personalization, one master tool, ideas for novel tools* |
| Hardware | *Sticky notes, paper, tagging, cloud, phone camera, phone dictation* |
| Archives | *Revisiting ideas, naming conventions, idea bank, inspirational materials, finding ideas, folder organization, forgotten ideas, desk area* |
| Collective ideation | *Decision making processes, ideation in a company, collaborating with a whiteboard, other tools for collaboration* |
| Personal ideation | *Ideation process, markers/tags to self* |
| Representing ideas | *Challenge of representing ideas, communicating ideas, flow of ideas* |

In the following axial coding, we moved towards *analytic codes* (Gibbs 2018), which identify and describe the central activities around the core phenomenon idea management. These codes are presented in the findings as a comprehensive model of idea management. From this model, the category 'strategies' (actions taken in response to the common phenomenon 'idea management') (Creswell 2013) is the basis of the framework of strategies for idea management presented in this paper.

The codes and categories described in the findings are derived from a combination of reflection on vocabulary used by the interview participants themselves, and concepts from PIM, creativity theory, and interaction design theory. Because the goal of the framework was for it to resonate with both practitioners and researchers, the terminology reflects both practice and theory.

## 4 FINDINGS

In this section, we will present first a grounded theory-model describing the core phenomenon of 'idea management' exemplified with quotes from the participants in our studies. This model is based on axial coding of the interview data, and divided into categories: *causal conditions* (factors which cause the central phenomenon), *central phenomenon* (a process or action, in this case *idea management*), *strategies* (actions taken in response to the core phenomenon), *contextual and intervening conditions* (situational factors, which influence the strategies), and *consequences* (outcomes from using the strategies) (Creswell 2013).

Secondly, we will present the other main contribution of this paper: a framework of strategies of idea management - that is, strategies that interaction designers take in response to the core phenomenon of managing ideas. In line with previous work (Kidd 1994; Shneiderman 2007; Bernal et al. 2015, Maudet 2017), we argue that the development of creativity support tools would benefit from an increased focus on supporting the *creator*, rather than on the created product. In this vein it is important to understand not only *what* the designer does, but also *why* they do it. Our framework provides a detailed description of the explicit or tacit reasoning behind the activities involved in idea management in the frame of professional interaction design.

## 4.1 A grounded theory-model of idea management

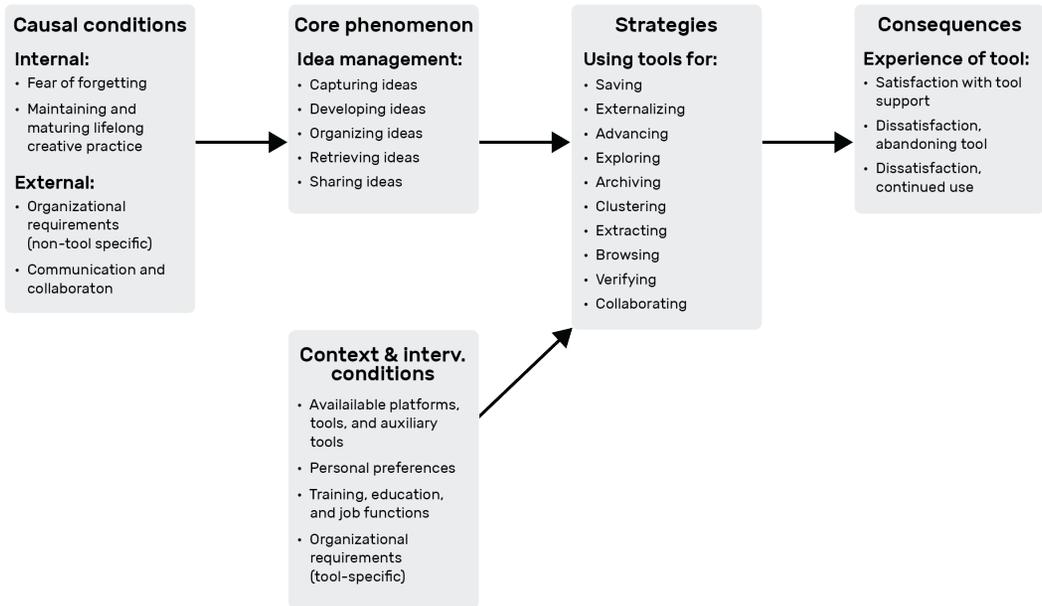

Fig 2. A grounded theory of professional idea management.

The theory of the core phenomenon and surrounding factors is shown in figure 2. The goal of a grounded theory-analysis is to understand a process or an action involving several individuals, and in this case several tools (Creswell 2013). This theory thus explains how professional idea management is performed and supported over time, by tools, through actions and strategies taken by individuals, who experience individual outcomes. The contents of the model are described in detail in the following sections.

*4.1.1    Core phenomenon.* Figure 3 describes the activities involved in the *core phenomenon*, i.e. idea management. This phenomenon consists of activities of capturing, developing, organizing, retrieving, and sharing ideas. These activities correspond to descriptions of idea- and information management presented in earlier work (Barreau and Nardi 1995; Coughlan and Johnson 2008; Efimova 2009).

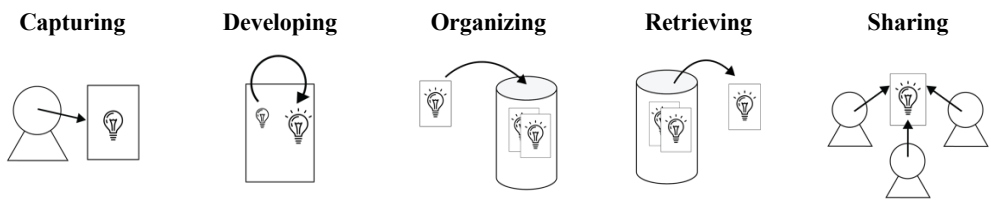

Fig. 3. Central activities of idea management. First, an idea is **captured** or externalized by a designer into the world. Externalized ideas can be **developed** and manipulated. They are usually **organized** in a repository, from which they can be **retrieved** at any time. Finally, ideas and externalizations can be **shared** with others. Aside from capture, which precedes further engagement with the idea, the activities do not happen in any particular chronological order, but can supersede each other in any sequence.

*Capturing* ideas equals externalizing the idea or a representation of the idea in an analog or digital form - the active shaping of the world as an intellectual resource (Dix and Gongora 2011). This can happen during design work or at other places and times: "*Actually I was at lunch the other*

*day, sometimes I use my mobile phone, but those are just for note taking purposes. I don't draw, I don't have any tools I would use to sketch an idea. But I didn't have access to my mobile phone, so I used the back of the receipt. (...) I mapped out an entire solution on the back of a receipt. (...) it was exactly what I needed at that moment"* (P9).

*Developing* ideas means expanding on the idea, whether that is contradicting it, reshaping it, or advancing it. It is a part of idea management, as the format of the idea often changes during this process - maybe the idea "moves" from a note on a receipt to a digital format. One designer (P5) described using a separate email account to manage his ideas, where each email thread represented a separate idea. If he pressed 'reply' to a thread, it meant he was contradicting the central idea, and if he pressed 'forward', it meant he was expanding on the idea.

*Organizing* ideas includes filing, storing, and rearranging ideas in a systematic fashion. Most of the study participants had developed an individual system for identifying ideas in progress (*working* ideas) from ideas that they were no longer working on (*archived* ideas): *"...my structure of my files is that I have a version 1, version 2, version 3. So, I never delete anything. But I just put the latest file on the outside of my folder. And then that way I can always look back on my ideas and have a folder for inspiration"* (P8). These systems would rely heavily on the individual designer and their memory, and do not always work for shared depositories of files.

*Retrieving* ideas equals finding stored ideas again. This can happen incidentally by browsing: "*Every once in a while, I will find one [a notebook], and I'll flip through it, and there's some interesting things there. There's this spark of something that actually came out and was handed off to someone else on the team*" (P9) or deliberately by searching "*I will go back to old ideas. One, for reference of the question of what did we do and why? The rationale of something, I guess it's sort of a portrayal. (...) The limitation is my phone space really. The times I go back is when there's a question about why we did something, the way we did it or if we needed to go over the rationale for something*" (P6).

*Sharing* ideas involves any activity where a person other than the creator of an idea is included, for instance during collaborative design sessions. The externalization of the idea usually serves an informational purpose (Dix and Gongora 2011), and the representation of the idea is central: "*It can be difficult for me to understand when people say 'We'll just do this!' and then they make wireframes on boards, because it doesn't speak to me. An ipsum-text and some squares, it doesn't appeal to me. It's also because I'm so visually oriented, I need something more than just squares with X's. I always get upset if someone has designed something with ipsum, because then you can make everything look nice. Our titles will often be super long because it's the users who write them*" (P18).

*4.1.2    Causal conditions.* A priori, the need for using tools to manage ideas stems from the limited cognitive abilities of the human mind (Scaife and Rogers 2005). But fear of forgetting ideas is only part of the story. Previous research has demonstrated how information management is influenced by value goals as much as resource accessibility (Kaye et al. 2006), and we confirmed this in our analysis.

There are external and internal causal conditions of professional idea management. This distinction is related to the notion of intrinsic versus extrinsic motivation of creative work: intrinsic motivation arises from the intrinsic value of the work for the individual themselves, and extrinsic motivation arises from the desire to obtain outcomes distinct from the work itself (Amabile 1993).

The interaction designers we interviewed described the following **intrinsic motivational factors** for using tools to manage ideas:

- **Fear of forgetting or losing the idea**: *"I forgot what that meant [points to entry in Google Keep] (...) I just wrote it down and I don't know what it was for. I'll probably archive it later, but I'll probably keep it there, just in case I might need it"* (P8).

- **Maintaining and maturing a lifelong creative career:** As shown by Kaye et al. (2006), archiving and managing ideas occurs for value-purposes as well as functional ones. For interaction designers, ideas and the representation of them are some of the most important materials they work with in a professional context. Depending on the individual designer, the use of idea archives has been shown to range from a problem-solving focus, where finished products and their working files inform the designer in similar cases, to an artisan design-focus, where ideas are explored and developed over extended periods of time with no particular goal in mind (Inie et al., 2018b). In the interviews of this paper, a majority of the interaction designers expressed the need to externalize and preserve their ideas for creative purposes beyond organizational due diligence: "**Why do you capture ideas?** *I've never tried to forget an idea, so that's not the reason. I think it becomes tangible when you write something down, it's just like goals writing for yourself, it becomes tangible. You have to take like, "Okay, but that's not even possible, so how would I make it possible?" So, you're obliged, you're making it something real. Everything starts with the thought, and then when you put it here it becomes a bit more tangible, and then (...) "Oh my God, they implemented it." And then, "Oh my God a service technician is using it!"* (P15).

Additionally, the designers were motivated by the following **external conditions**:

- **Organizational requirements (non-tool specific).** Few companies require designers to share their ideas specifically, but most companies expect designers to store and share their working files, i.e. representations of ideas.

- **Communication and collaboration.** For the purpose of sharing ideas or continuously collaborating with others around the development of ideas, it is necessary for designers to be able to refer others to a reliable storage source of representations of ideas.

*4.1.3 Contextual and intervening conditions.* In our analysis, the relationship between the category 'causal conditions' and the category 'contextual and intervening conditions' is that 'causal conditions' describe the factors that lead to the activities involved in idea management taking place, and 'contextual and intervening conditions' are the factors influencing how the individual strategies of idea management are chosen and performed. We identified the following contextual and intervening conditions of tool use in professional idea management:

- **Available platforms, tools, and auxiliary tools.** What tools are available depends not only on what tools exist, but on which tools the designer knows about. Many of the study participants asked us to share the results of the study because they were curious to learn about the tools others use, that they did not know of. One designer said: *"If I get hold of a really bad, pointy, stiff pen on a bad writing surface, right? It doesn't conceive any ideas in me whatsoever. If you get the right marker which glides in the right way on a whiteboard, you know, and makes a thick, fat line - that's what you need, right? That raises your productivity and your, like, creative desire. It's a total turnoff to get a ridiculous little pointy pencil you can only draw little crosses and checkmarks with, right? I think, like a musical instrument, if you give people a wrong violin, then - they can play it, but it won't be great art"* (P4).

- **Personal preferences.** Which tools the designers preferred to work with depended, to a large extent, on how often they captured ideas, which representation the tools allowed them, and how the designer planned to utilize the idea later, a similar finding to previous studies of information management (Boardman and Sasse 2004). We have previously published an analysis of different strategies for utilizing archived ideas (*reference anonymized for peer review*).

- **Training, education and job function.** "Interaction designer" is a job description encompassing many different design tasks, and often employing people from different educational backgrounds. For example, we interviewed a designer with a background in sociology, and another with a background in computer science. Someone who is trained to use specific software tends to prefer the representations and auxiliary tools that this software allows. We saw, for instance, that only the designers who had trained as *graphic* designers used Adobe Illustrator to capture or develop ideas. This is an example of how the tool shapes the human action as a co-adaptive phenomenon (Mackay 1990).

- **Organizational requirements (tool-specific).** The difference between non-tool specific organizational requirements and tool-specific requirements is whether the requirements are causal for idea management taking place, or whether they are intervening in the specific choice of tools. Both can be the case. Many professional designers are required by their workplace to use specific tools for e.g. security reasons: *"I think that's something worth mentioning, um, security does change the way…like the things that we can use. We used to be big Evernote people and then it kind of just […] They can't be put under our security. It's kind of like, things are confidential, and there's always, like, hackers that can get in so there's some things that we have to do, like, use sparingly, or use under the table and just use really smartly even though the tool itself is really helpful."* (P11).

*4.1.4 Strategies.* Strategies describe the actions taken in response to the core phenomenon (Creswell 2013). A strategy is a high-level plan for how to obtain certain goals, in this case the goal of managing ideas with the use of one or more tools. Therefore, it is central to identify the *goals* of managing ideas to be able to explain and define the strategies for doing so.

As an example, the action to *capture* an idea, as well as the tool chosen to assist this action, is highly influenced by the purpose of the capture. If the purpose of capture is to *retain* the idea, rather than to refer to it later, this purpose affects the tool choice: *"**Do you normally go back and look at [your notebook] again?** No, I have so many of these notebooks. And often times it's more for me as a tool to just write it for retention versus recall. I don't use it as much for recall"* (P9).

Distinguishing between different strategies can inform and inspire the development of tools that support the designer in achieving an underlying goal or objective. Therefore, our main contribution of this paper is the identification of a framework of strategies performed by interaction designers to manage ideas professionally: *saving, externalizing, advancing, exploring, archiving, clustering, extracting, browsing, verifying,* and *collaborating*. We will explain this framework and the goals of these strategies in more detail in a separate section of the findings.

*4.1.5 Consequences.* The consequences or outcomes of the strategies correspond to the interaction designer's experience of using a specific tool.

- **Satisfaction with tool support.** If the interaction designer finds themselves satisfied with the tool's support of the strategy they have chosen, they will most likely continue using the tool. The interaction designer may also expand their use of this tool to support other strategies than originally intended: *"I started a couple of years ago to use an email account to capture my ideas because, then I had a Blackberry with a keyboard, which I already liked, […] and then I started to think about "hey, why don't I just make a photo of my sketches and send them to this account too", and so I started to use email. I wasn't a conscious goal. It wasn't a direct decision"* (P5).

- **Dissatisfaction, abandoning tool.** If the interaction designer is dissatisfied with the tool's support of the strategy, they might abandon the tool, and may or may not find a new tool or return to a known one: *"I tried several tools. I tried pen and paper, I tried whiteboards, I tried it with Asana, I tried it with [company name], I tried several tools and,*

*in the end, I was enthusiastic first moment, and after a couple of weeks I recognized that I already stopped to use all these tools. And I thought about a tool which is present everywhere and which is easy to use and which I can access from everywhere […] and I recognized that email could do this stuff for me"* (P5).

- **Dissatisfaction, continued use.** If the interaction designer is not overwhelmingly dissatisfied with the tool, or if they are prevented by lack of time, lack of available tools, or organizational support, from finding a different tool, they may continue to use the tool. They may experience it as a pain point in their professional practice: *"We had so many problems with Asana. (…) The guy, Lucas, was just sitting three meters away from me, and I was like "Hey, come on - talk to me!". (…) I used it and he was really satisfied with it because he could track every process and look what happened then and 'are we on time with this task and can we postpone this task'. (…) he felt my reluctance, but I couldn't express why. (…) If I had to do collaboration and all that stuff and document or something like this, I used Asana, I put the document in this Asana box and said "Okay, hello there! I've put it in Asana". Or I just sent it via email. Lucas asked me to use Asana"* (P5).

## 4.2 A framework of strategies for idea management

A strategy can be understood as a high-level plan with the purpose of achieving one or more goals under conditions of uncertainty. It is central to discuss the strategies and goals of the activities in more detail.

Different tools are used for capture because the goal of capture varies. Consider how this designer uses different modalities of the platform "pen and paper" for different purposes:

*"**What's the difference between the graph paper and the notebook - how do you use those differently?** Notebook I take with me to meetings and I'm, just, general note taking. The notepad is more of actually sitting down and doing design work. (…) In the notebook that might be just, kind of… Just sometimes, you know, you just take notes just to highlight some things that you want to remember. Deliverables. You know, action items, stuff I need to, like, expand upon further later on or just some stuff you want to keep in your mind and that, it helps to just write down to remember better"*. (P3).

Because the goal of "not forgetting deliverables" is different from the goal of "actually doing design work", strategy and tool choice varies. The activity is still *capturing ideas,* but there are different expressions or strategies for capture. In this section we present a detailed description of each of the strategies we identified in our analysis: *saving, externalizing, advancing, exploring, archiving, clustering, extracting, browsing, verifying,* and *collaborating.*

*4.2.1 Saving ideas.* When the interaction designer *saves* ideas, they capture something with the intent of returning, or being able to return, to it later. A widely used example of this is taking a picture of a whiteboard that contains representations of ideas. As a specific example, designer P2 had built an entire digital database for 'long-term ideas' - ideas that he wished to be able to return to and develop at any given moment in the near or far future. These ideas were captured purely in text, so he could use an NLP algorithm to search for keywords that were vaguely related to those in the actual text. Other examples of saving strategies are screenshots on a phone or laptop, sometimes with accompanying annotations: *"If I see a great UI idea, I screen dump it and edit it with the pen tool. Sometimes it is, like, completely rough, and I don't think it makes any sense for anyone else but me"* (P18). See Fig. 4 for examples of this designer's screen shots with annotations. The ideals of the *saving* strategy were a combination of the capture process being quick and the tool readily available, as well as the representation providing exactly enough information for the capture to make sense when encountered later.

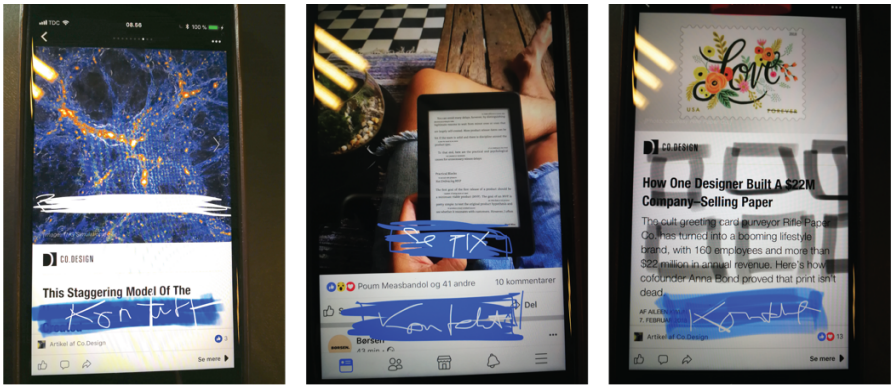

Fig. 4. Designer P18 'saving' captures.

*4.2.2 Externalizing ideas.* By *externalizing*, we mean initial capture of ideas which serve informational, formational, transformational, or transcendental purposes, as described by Dix and Gongora (2011). While Dix and Gongora describe these terms in much more extensive detail, we will describe this strategy as any process of making internal processes (e.g. ideas) external (Ibid.). The main difference between this strategy in relation to the strategy of *saving*, is that the designer does not necessarily intend to return to the externalization again, but rather the process of externalizing is the goal in itself. In the words of Kidd (1994) "*We may have been fooled into thinking that knowledge workers write things down because they need an external memory store, whereas in many cases, it may be the graphological act itself which is important*". Externalizing ideas does not only entail writing or sketching. In fact, several of the designers in our study mentioned deliberately using their own voice as a mode of externalizing: *"Typically, I'll maybe sit in the car and reflect over something for 30 minutes, and during that half hour, maybe four or five things will emerge, which actually are like 'Oops, there was something. That, I need to be able to return to"* (P7, authors' translation).

Externalizing allows the designer to release some of their working memory (Kirsh 2009). Very often, the tools used for externalizing would be an analog platform - either some form of pen and paper, or a whiteboard: *"And those [sticky notes] will then be translated at some point but I'm constantly using… It has to be pen to paper first for my ideas to get flushed out"* (P9). Several of the interaction designers in our study expressed that externalizing their thoughts in situ allowed them to remember something better without relying on the physical manifestation again. The ideal tools for the *externalizing* strategy were always described as having the least constraining properties possible, and often being analog: *"As soon as you start making it ultra-concrete in [digital] wireframes, a lot of things emerge that one had not seen before. So, a development happens there. But I'd say the foundation is laid on paper"* (P7).

*4.2.3 Advancing ideas.* Advancing the idea is a strategy for the idea management activity of *developing* ideas. When the designer is *advancing*, they are working more convergently than divergently - working towards solving a specific problem or refining a solution to a given problem. It may, for instance, be the process of transforming an analog sketch into a digital format, or the process of refining an existing design artifact. While this development is happening, new ideas can still emerge and old ones can be abandoned; however, that is not the primary goal of the strategy. *Advancing* of ideas was often described by our study participants as "*actual* design work", which of course includes many processes of trial and error, conceptual detours, and discoveries of new properties. The primary goal of this strategy is, however, to bring a design idea closer to a finished artifact. Because our study subjects were in the business of interaction design, the tools involved in this strategy are usually digital. The ideals of tools for this strategy vary greatly with the individual designer's personal preferences, background, skills, and workflow. Because the designer often has a more or less defined end-goal in mind, efficiency of moving towards that goal is an essential property: *"**Imagine an ideal tool in your mind (...)**. So, it would be something maybe with VR because then I could just ... Okay, now I'm really out there. But something where I could actually draw when I was standing here, so I'm interacting with the [artifact], I'm building screen by screen and I'm not, again, caught into a tablet. (...) And then I could just, like, "Okay, so I press this button and communicate." And then it would already know how the communication protocols between the [artifact] and this would work. So, it would just know, "Okay, so this would flow back and forth and this is how we get the right parameters here"* (P15).

*4.2.4  Exploring ideas.* This strategy, under the activity of *developing* ideas, refers to when the designer divergently and open-endedly explores an existing idea or its properties. The difference between *exploring* and the strategy *advancing* is the thinking involved: where the designer who is *advancing* an idea is focused on the convergence and the efficacy that the tool in use allows, the designer who is *exploring* an idea is thinking primarily divergently, open to creative detours and novel visualizations. In fact, one desired feature often mentioned by the participants was for a given tool to provide different visualizations of a working idea: *"I need a place to sketch things out. Something that doesn't feel limited--something that I can have as much space as I need to keep growing my ideas, keep growing, like brainstorming. Of course, written text is always helpful as well, but I would say that if there could be a good balance between images, pictures, and text, that would be really nice"* (P13). While some of the study participants often described the *advancing* strategy as "*actual* design work", it is our impression that they considered the *exploring* strategy to be more tightly connected to the core concept of creativity. Inie et al. (2018b) explores the difference between an *artisan design* and a *problem solving*-approach for using idea archives in day-to-day design work, and highlights that both are central capabilities for a professional interaction designer.

*4.2.5  Archiving ideas. Archiving* describes the archival of ideas in a digital or analog storage facility, either deliberately or automatically. An example of an automatic archiving system is the synchronization of a smartphone gallery to a cloud folder. The paramount property of this strategy is for the designer to feel confident in leaving the manifestation of the idea and trust that it will stay in this storage and be accessible if it becomes relevant in the future. The study participants described relying heavily on cloud storage for both personal and shared archives, but also on physical space to some extent: *"I write a lot of UI microcopy for work, so I always keep, like, certain things in it, like a certain area in front of my computer. Like, the rules that I follow for my microcopy are always right there (...) and those never go away"* (P1), and *"I have this filing cabinet next to my desk, where I keep this. I keep all my printouts of any time I printed something out for a project"* (P11).

*4.2.6  Clustering ideas.* After different ideas have been externalized, the interaction designer may spend time *clustering* or applying systems to the externalizations. Sticky notes are particularly popular for this purpose, but clustering may also involve going back to old archives and reorganizing files. When we asked designers to imagine novel tools for idea management, a very commonly expressed ideal was some form of a large, interactive interface which would allow them to consolidate and group various media forms: *"if you had an invisible, I guess, interface, which would be great to have. Some sort of retinal display where you could move ideas around. So actually, it wasn't a computer, it wasn't a device, it was part of a biometric interface...Bringing it back to today, I really didn't love Google, Google Drive. (...) If you're not familiar with the structure, if you don't have an idea of what filing system you're going to use, then it can actually be pretty daunting because you start from somewhere and it becomes a real mess really quickly because you have lots of files without categorization file folders or structure. (...) I would love that intelligent interface to file my documents and thoughts without me having to think about it, so it'd be based on*

*the content in there or the type of idea that I'm coming up with. (...) I spend a lot of time trying to order my thoughts"* (P10).

*4.2.7    Extracting ideas.* Under the activity of retrieval of ideas, the most commonly mentioned reason for returning to old ideas was to try to discern the rationale of previous design processes. When *extracting*, the designer is trying to find or identify a *specific* idea and the rationale or decisions surrounding it. The search itself can be performed by both browsing and searching (Boardman and Sasse 2004), but the objective is to find something specific: *"I will go back to old ideas. One for reference of the question of what did we do and why? The rationale of something (...) The times I go back is when there's a question about why we did something, the way we did it or if we needed to go over the rationale for something. Or if someone was held accountable for something and we don't know who to hold accountable for it"* (P6).

*4.2.8    Browsing ideas.* The search strategy of *browsing* describes the process when the designer looks through old ideas without a particular goal in mind. 'Old ideas' can entail both ideas that the designers may have externalized themselves, or representations of others' ideas that the designer has previously saved: "*I took a picture of it and then uploaded it to Evernote for later usage. It wasn't something that has anything to do with what I'm working on right now. But I found that adding it to this note is a really good idea because I might later on touch on something that has a part of it in it or in some way it will connect things for me when I get tasks later on. So, it's very useful to have this [...] library of things that inspire me for whatever reason (...) So when I'm stuck or I need inspiration in some way, I'll go into this...It's basically this image base of things I've taken a picture of in some way, and I'll just look through it randomly"* (P12). A desired property of tools that support this strategy was the option to annotate, tag, and comment on the ideas - so the designer would spend less energy remembering why the idea was saved in the first place. This finding is parallel to the findings of previous studies, which have shown that designers wish to know the *stories* behind the saved artifact, rather than a "disembodied" representation of it (Herring et al. 2009; Sharmin et al. 2009).

*4.2.9    Verifying ideas.* Sharing ideas with the purpose of *verifying* them is different from actively collaborating on them. The interaction designers of our study described using slideshow software for collaboration purposes, but when we inquired further, it was very clear that slides were used to *present* ideas to other stakeholders, rather than to actively collaborate on generating or developing ideas. For tools supporting this strategy it is more critical to help communicate the idea correctly, than to allow for several people to contribute to the idea on the spot: *"When presenting ideas, I find it most effective using a PowerPoint or a Keynote. The reason why is because people can only intake so much information, like if I showed them the entire Google Doc or something, they don't know (...). But then when I arrange slides I try to get to like the main thing. And go from slide to slide to slide. So, I just present bits and pieces of information so it's easier to digest"* (P8).

*4.2.10    Collaborating on ideas. Collaborating* on ideas describes several individuals working simultaneously together on generation and/or development of ideas. The quality of tools to work as mediation (Dalsgaard 2017) is important during such activities. The most important property mentioned by our study participants was the ability of a tool to provide a comprehensive overview, while allowing all participating members of the group to contribute to the shared pool of ideas: *"I would love a huge interactive touchscreen in my day where I could doodle, I could draw, I could swipe, I could write, I could pull up images from the net and having everything there at my fingertips. It's not just having a touch laptop. It's having a screen the size of a wall that you can be able to share or interact with many people at once. For overview and just ideation. I mean whiteboards and dry erase markers, I mean, sticky notes are great, but it's pretty inefficient and you're always referencing different things"* (P10).

An overview of the ten strategies is presented in table 5. The strategies unfold the activities described in figure 2 and section 4.1.1. The table shows how each activity of idea management can be performed as a strategy of primarily divergent or convergent thinking (Runco et al. 2006). Because interaction designers operate in scopes of both creative, open-ended discoveries and of concrete problem solving (Inie et al. 2018b), they employ both divergent and convergent strategies to pursue their goals. This duality is simplified as the expression of two different strategies within each idea management activity, for instance *saving* and *externalizing*, which are both strategies within the activity *capture*. When the designer is *saving* an idea, they are capturing something with relatively specific goals in mind: to not lose the idea again, and to be able to retrieve it in the future. When the designer is *externalizing*, they have a less specific goal in mind: they are exploring options, and potentially new ideas and problems.

Furthermore, Table 5 also lists the desired properties of tools to support these strategies as expressed by our study participants, as well as the tools and platforms which currently support them. We include the list of tools mentioned to give a detailed idea of how the strategies were described by the interaction designers who were part of the study in this paper.

Table 5. A framework of strategies for professional idea management.

| Activity | Strategy | Thinking | Desired properties of support tools | Primary tools and platforms |
| --- | --- | --- | --- | --- |
| Capturing | Saving | Convergent | Efficiency, availability | Phone camera, digital note taking software, pen and paper |
| | Externalizing | Divergent | Freedom of expression | Pen and paper, whiteboard |
| Developing | Advancing | Convergent | Efficacy, efficiency | Design and UX software, pen and paper, whiteboard |
| | Exploring | Divergent | Freedom of expression, different visualizations | Pen and paper, whiteboard |
| Organizing | Archiving | Convergent | Accessibility, reliability | Cloud storage services, computer folders, physical storage units, desktop |
| | Clustering | Divergent | Different visualizations | Sticky notes, computer folders |
| Retrieving | Extracting | Convergent | Accessibility, efficiency | Cloud storage services, computer folders, physical storage units, desktop |
| | Browsing | Divergent | Providing inspiration, showing the core of the idea | Cloud storage services, computer folders, physical storage units, desktop |
| Sharing | Verifying | Convergent | Correctly communicating the idea | Presentation software |
| | Collaborating | Divergent | Freedom of expression, providing overview | Whiteboard, pen and paper |

## 5 DISCUSSION

There are many challenges to studying tool-use and creativity in professional settings, and many ways to approach this complex subject. The creative objectives of many designers are detailed and complex, and looking at strategies in single use cases provides us with a better understanding of design practice. As an example, previous studies have shown a general preference in the activity *retrieving information* for browsing over searching (Boardman and Sasse 2004), but our data suggests that this depends on the purpose of retrieval. With this study, we have aimed towards gathering small-sample, deep knowledge and thereby establishing a rich foundation for understanding real-life practices. In this section we will discuss the implications of the work presented in this paper.

Firstly, the analysis **presents descriptive knowledge** about how interaction designers work. Even at a descriptive level, this knowledge is far from trivial, as it is inherently difficult to gather knowledge about idea management in organizational settings and outside the work environment (Coughlan and Johnson 2008).

Secondly, this paper **presents a theory of professional idea management.** The term 'idea management' is neither widely used nor well defined, but we argue that it is an area of great significance for the practicing interaction designer, and potentially also to a wider range of professions. Constructing, defining and discussing theories of processes is an essential basis for research, and thus the analysis in this paper advances a relatively novel research field.

Thirdly, this paper **offers a framework to assist researchers in analyzing, understanding, and describing professional tool-use**, based on strategies and goals of practicing interaction designers. This framework is useful for researchers as well as for developers of existing and novel creativity support tools in characterizing the role of tools and their support. Categorizing tools in terms of the goals they fulfill to interaction designers may also reveal which goals are *not* fulfilled using specific tools, or how designers modify tools to fit their purposes. For example, one designer (P18) had created a personal Slack channel with herself, which she used as a to do-list - because she would always open Slack as part of her work day. If we examine solely the functions of Slack, the tool is designed for collaboration purposes rather than for individual idea capture. But if we consider the desired property in a tool for *archiving* (accessibility and reliability), Slack fulfills these needs well. Designer (P5) utilized a separate email account as both a storage and development tool for his ideas. This is another use we might not expect from looking at the inherent functions of email as a tool. However, if we think of desired properties of a tool for *saving*, *advancing*, *archiving*, and *retrieving*, email is an appropriate tool to fulfill these goals.

Overall, this paper contributes to the discipline of research in professional interaction design. It is evident from the study that there is not one perfect tool for idea management, nor one standard process or system. Interaction designers create their own assemblages of tools, methods, and systems that work for them individually. Even though each idea management system described in the study was unique, each system was internally coherent. One designer would not use Evernote to keep track of ideas for one project and Google Docs for the next project, for instance. They would adhere to some system that fit their individual needs.

The model offers an explanation and analysis of idea management *processes*, which is central if we wish to be able to develop tools that are process-focused rather than product-focused. Practicing interaction designers do not always work with a high certainty of the goal they are trying to achieve. Generating and developing design ideas involves a high degree of open-ended exploration and experimentation, and currently-available tools do not always support this. Developing tools with a process focus might entail providing the interaction designer with open-ended, easily customizable interfaces, different visualizations, and continuously offering inspiration from external and archived resources of ideas.

Another example of how to use the framework is to deliberately facilitate one strategy at a time. We can aim to support the strategy *verification* by, for instance, developing tools to propose different visualizations or representations of the same file, depending on the stakeholder the designer is presenting to. Or, with the strategy of *saving*, by developing tools specialized for different forms of tagging and annotation, to allow for rapid capture as well as easy retrieval.

Further studies should unquestionably be performed to expand upon our findings. Analyzing idea management strategies within one team of designers might yield a different result than that from within another team, because the strategies are affected by context and intervening conditions, as shown in the grounded theory-model. However, the framework might inspire the adaptation of novel tools and novel processes for idea management.

## 5.1 Limitations and future work

The goal of this research was to generate knowledge for researchers and developers of creativity support tools to allow them to better understand and support real-world practices. One of the caveats of interviews is that they report on what study participant *say* they do, rather than their observed practices. Future studies of mixed methods would expand upon the findings with the generation of more and new insights that were out of the scope of this work.

We have used qualitative interviewing because of its strength as a means of accessing attitudes and values that would not otherwise have been observable or discoverable (Byrne 2004). There are major qualities of the methodology, among other the ability to generate new insights and hypotheses about causes and relationships and the appropriateness for studying organizational creativity with good ecological validity, that is, providing insights that are true to real-world practices (Amabile and Mueller 2008). As with any research method, there are also caveats. The limitations of this small-sample work are that it does not allow for determining causal influences on practices, for testing of hypotheses about relationships between causal factors to be tested, or for broad generalizations (Ibid.).

Future research could expand upon the findings by combining qualitative interviews with, for instance:

**Longitudinal observations in the workplace.** Longitudinal observations of interaction design practices would offer additional deep knowledge of processes for idea creation, management and sharing. This type of study is promising for unfolding some of the findings presented in this paper, such as the context and intervening conditions that influence the chosen strategies.

**Screen capture and video recordings** of activities happening at the desktop in real time. To gather more detailed knowledge of tool-use, it would be extremely interesting to have access to screen captures or video recordings of workstations over an extended period of time. Such studies would be less intrusive to the normal work practice, and could help discover tacit patterns or tool-uses. Ideally, such recordings would be combined with following interviews that allowed the researcher to ask clarifying questions. This type of study would be particularly interesting for discovering the correspondence between what interaction designers say they do and what they actually do. It could be a very promising avenue of research in terms of watching idea development over time, both on a short- and long-term scale.

**Building and testing tools with a process focus.** As described in the introduction, there is potential in developing tools with a process- as well as a product focus; that is, tools which support creative strategies in an unobtrusive manner. Future work would ideally be directed towards testing novel tools, and exploring how these might be beneficially adapted in professional interaction design practices. The framework of strategies for professional idea management we have presented in this article may inform the development of novel tools.

## 6 CONCLUSIONS

This work presents an extensive analysis of the activities and processes involved in professional, tool-supported idea management. We suggest that this analysis can benefit research into information management and creativity support tools by providing a deep, detailed understanding of real-world practices. Our findings allow us to characterize the idea management process in terms of relationships between the core phenomenon and its activities, causal conditions, context and intervening conditions, strategies taken in response to the core phenomenon, and consequences of using the strategies.

We identified a unifying theory of the core phenomenon of idea management: capture, development, organization, and sharing of ideas. We presented an analysis of this phenomenon as influenced by internal- (fear of forgetting, and maintaining and maturing a lifelong, creative practice) and external- (organizational, non-tool specific requirements, communication, and collaboration) causal conditions. We identified the context and intervening conditions of idea management as available platforms, tools, and auxiliary tools, personal preferences, training and job functions, and organizational (tool-specific) requirements. Our primary contribution of the theory was the identification of ten strategies, which are actions taken in response to the core phenomenon. These strategies were *saving, externalizing, advancing, exploring, archiving, clustering, verifying,* and *collaborating*. And lastly, we described the consequences of the strategies taken as either satisfaction with the support of the tool utilized, dissatisfaction and abandoning of the tool, or dissatisfaction, but continuation of using the tool.

Finally, we elaborated on the strategies, and illustrated them as a framework which might be used to analyze and describe specific tools based on how they fulfill certain strategies or objectives to the designer, rather than the tools' inherent functionality. With this work, we hope to inspire both developers of novel creativity support tools as well as researchers in interaction design.

## ACKNOWLEDGMENTS

This work was funded by the CIBIS (Creativity In Blended Interaction Spaces) grant from the Innovation Fund Denmark and by Digital Tools in Collaborative Creativity (CoCreate) from the Velux Foundation.